\documentclass[hidelinks,10pt,letterpaper]{article}

\usepackage{helvet}
\usepackage{courier}
\usepackage{epsf}
\usepackage{epsfig}
\usepackage{mathrsfs}
\usepackage{subfigure}
\usepackage{graphics}
\usepackage{balance}
\usepackage{graphicx}
\usepackage{balance}
\usepackage{epstopdf}
\usepackage{pifont}

\usepackage{amsfonts}

\usepackage{pifont}

\usepackage{epstopdf}
\usepackage{etoolbox}
\newcommand{\method}{{\tt LinNet}}
\newcommand{\brier}{\beta}

\usepackage[top=0.85in,left=2.75in,footskip=0.75in]{geometry}
\usepackage{bm}
\usepackage{changepage}

\usepackage[utf8]{inputenc}

\usepackage{textcomp,marvosym}

\usepackage{fixltx2e}

\usepackage{amsmath,amssymb}

\usepackage{cite}

\usepackage{nameref,hyperref}

\usepackage[right]{lineno}

\usepackage{mdframed}

\usepackage{microtype}
\DisableLigatures[f]{encoding = *, family = * }

\usepackage{rotating}
\usepackage{amsmath}

\newcommand{\weight}{w}

\raggedright
\usepackage[aboveskip=1pt,labelfont=bf,labelsep=period,justification=raggedright,singlelinecheck=off]{caption}

\makeatletter
\renewcommand{\@biblabel}[1]{\quad#1.}
\makeatother

\date{}

\usepackage{lastpage,fancyhdr,graphicx}
\usepackage{booktabs}
 \usepackage{multirow}
\fancyheadoffset[L]{2.25in}
\fancyfootoffset[L]{2.25in}
\usepackage{tcolorbox}
\tcbuselibrary{theorems}

\newtcbtheorem[number within=section]{mydefinition}{Definition}%
{colback=yellow!5,colframe=yellow!35!black,fonttitle=\bfseries}{th}

\newcommand{\rating}{{r}}
\newcommand{\lineup}{{\lambda}}
\newcommand{\nodes}{\mathcal{V}}
\newcommand{\edges}{\mathcal{E}}
\newcommand{\weights}{\mathcal{W}}
\newcommand{\lspace}{\mathcal{X}}
\newcommand{\lfeature}{\mathbf{x}}
\newcommand{\simil}{{\sigma}}
\newcommand{\edge}{{e}}
\newcommand{\node}{{u}}
\newcommand{\net}{\mathcal{G}}

\begin{document}
\vspace*{0.35in}

\begin{flushleft}
{\bf {\method}: Probabilistic Lineup Evaluation Through Network Embedding}
\newline
\begin{center}
\end{center}
\vspace{-0.5in}
\center{
Konstantinos Pelechrinis
\\
School of Computing and Information \\
University of Pittsburgh
\\

%
%





kpele@pitt.edu
}
\end{flushleft}

\begin{abstract}
Which of your team's possible lineups has the best chances against each of your opponents possible lineups? 
In order to answer this question we develop {\method}. 
{\method} exploits the dynamics of a directed network that captures the performance of lineups at their matchups.   
The nodes of this network represent the different lineups, while an edge from node $j$ to node $i$ exists if lineup $\lineup_i$ has outperformed lineup $\lineup_j$. 
We further annotate each edge with the corresponding performance margin (point margin per minute). 
We then utilize this structure to learn a set of latent features for each node (i.e., lineup) using the {\bf node2vec} framework. 
Consequently, {\method} builds a model on this latent space for the probability of lineup $\lineup_A$ beating lineup $\lineup_B$. 
We evaluate {\method} using NBA lineup data from the five seasons between 2007-08 and 2011-12. 
Our results indicate that our method has an out-of-sample accuracy of 67\%.  
In comparison, utilizing the adjusted plus-minus of the players within a lineup for the same prediction problem provides an accuracy of 55\%.  
More importantly, the probabilities are well-calibrated as shown by the probability validation curves. 
One of the benefits of {\method} - apart from its accuracy - is that it is generic and can be applied in different sports since the only input required is the lineups' matchup performances, i.e., not sport-specific features are needed.
\end{abstract}
 
\section{Introduction}
\label{sec:intro}

One of the decisions that a basketball coach has to make constantly is what lineup to play in order to maximize the probability of outperforming the opponent's lineup currently on the court. 
This lineup evaluation problem has been traditionally addressed through player and lineup ratings based on (adjusted) plus/minus-like approaches.  
In this work, we propose a different approach that is based on network science methods.  
In particular, we first define the matchup network: 

\begin{mydefinition}{Matchup Network}{theoexample}
The matchup network $\net=(\nodes,\edges,\weights)$, is a weighted directed network where nodes represent lineups. An edge $\edge_{i,j} \in \edges$ points from node $i\in \nodes$ to node $j\in \nodes$ iff lineup $j$ has outperformed lineup $i$.  The edge weight $\weight_{\edge_{i,j}}$ is equal to the performance margin of the corresponding matchup.
\end{mydefinition}

\begin{figure}[h]
\includegraphics[scale=0.4]{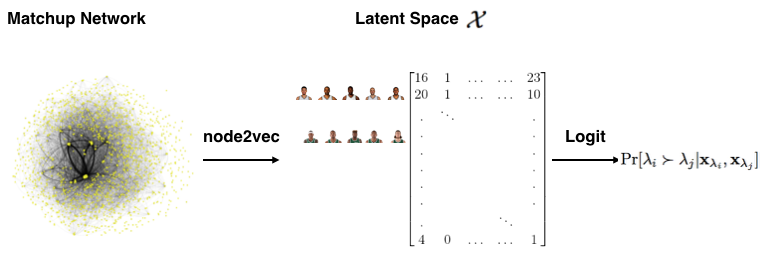}
\caption{{\bf The {\method} lineup evaluation method}}
\label{fig:ballnet}
\end{figure}

Using this network we then obtain a network embedding, which projects the network nodes on a latent space $\lspace$. 
For our purposes we adopt the {\bf node2vec} \cite{grover2016node2vec} framework for learning the latent space. 
Simply put the embedding learns a set of features $\lfeature_{\node}$ for node {\node}.  
These features are then utilized to build a logistic regression model for the probability of lineup $\lineup_i$ outperforming lineup $\lineup_j$, $\Pr[\lineup_i \succ \lineup_j | \lfeature_{\lineup_i},\lfeature_{\lineup_j}]$. 
Figure \ref{fig:ballnet} visually captures {\method}. 

Our evaluations indicate that {\method} is able to predict the outcome of a lineup matchup correctly with 67\% accuracy, while the probabilities are well-calibrated with a Brier score of 0.19. 
Furthermore, the probability validation curve of {\method} is statistically indistinguishable from the $y=x$ line and hence, the logistic regression model captures accurately the lineup's matchup probabilities. 
In comparison, we evaluate two baseline methods; (i) a PageRank-based ranking using the same matchup lineup networks, and (ii) a model based on the adjusted plus/minus of the players consisting each lineup. 
These two methods have accuracy ranging between 52-58\%. 

The rest of the paper is organized as following. 
In Section \ref{sec:materials} we present in details the operations of {\method} as well as the datasets we used. 
Section \ref{sec:analysis} presents our results, while Section \ref{sec:discussion} discusses the implications and limitations of our work.   

\section{Materials and Methods}
\label{sec:materials}

In this section we will present in detail (a) the design of {\method}, (b) the baseline methods for comparison, and (c) the datasets we used for our evaluations. 

\subsection{{\method}}
\label{sec:linnet}

The first step of {\method} is defining the matchup network $\net$. 
There is flexibility in choosing the performance margin that one can use for the edge weights. 
In the current implementation of {\method}, the weights of $\net$ correspond to the point margin per minute for the two lineups. 

Once the network is obtained the next step is to learn the network embedding. 
As our network embedding mechanism we will utilize the approach proposed by Grover and Leskovec \cite{grover2016node2vec}, namely, node2vec. 
node2vec utilizes (2$^{nd}$ order) random walks on the network in order to learn the latent features of the nodes, i.e., a function $f : \nodes \rightarrow \Re^d$, where $d$ is the dimensionality of the latent space.  
Starting from node $\node$ in the network and following the random walk strategy $R$ the network neighborhood $N_R(\node)$ of $\node$ is defined. 
Then node2vec learns the network embedding $f$ by solving the following optimization problem:

\begin{equation}
\max_{f} \sum_{\node \in \nodes} \log(\Pr[N_R(\node)|f(\node)])
\label{eq:opt}
\end{equation} 

Simply put, the network embedding maximizes the log-likelihood of observing a network neighborhood $N_R(\node)$ for node $\node$ conditioned on the network embedding $f$. 
The random walk strategy is defined by two parameters, $p$ and $q$, that offer a balance between a purely breadth-first search walk and a purely depth-first search walk.  
In particular, the random walk strategy of node2vec includes a bias term $\alpha$ controlled by parameters p and q.  
Assuming that a random walk is on node $\node$ (coming from node $v$), the unnormalized transition probability $\pi_{\node x} = \alpha_{pq}(v,x)\cdot \weight_{\node x}$. 
With $d_{\node x}$ being the shortest path distance between $\node$ and $x$ we have:

 \[   
\pi_{\node x} = 
     \begin{cases}
      1/p &,~ if~ d_{\node x}=0\\
       1 &,~if~ d_{\node x}=1 \\
       1/q &,~ if~ d_{\node x} = 2 \\ 
     \end{cases}
\]

As alluded to above parameters $p$ and $q$ control the type of network neighborhood $N_R(\node)$ we obtain. 
Different sampling strategies will provide different embeddings. 
For example, if we are interested in having set of nodes that are tightly connected in the original network close to each other in the latent space, $p$ and $q$ need to be picked in such a way that allows for ``local'' sampling. 
In our application we are interested more in identifying structurally equivalent nodes, i.e., nodes that are similar because of their connections in the network are similar (not necessarily close to each other with respect to network distance). 
This requires a sampling strategy that allows for the network neighborhood of a node to include nodes that are further away as well. 
Given this objective and the recommendations by Grover and Leskovec \cite{grover2016node2vec} we choose $q=3$ and $p=0.5$ for our evaluations.  
Furthermore, we generate 3,000 walks for each network, of 3,500 hops each. 
Finally, we choose as our latent space dimensionality, $d = 128$. 
Increasing the dimensionality of the space improves the quality of the embedding as one might have expected, however, our experiments indicate that increasing further the dimensionality beyond $d=128$ we operate with diminishing returns (with regards to computational cost and improvement in performance). 

Once the latent space $\lspace$ is obtained, we can build a logistic regression model for the probability of lineup $\lineup_i$ outperforming $\lineup_j$. 
In particular, we use the Bradley-Terry model. 
The Bradley-Terry model is a method for ordering a given set of items based on their characteristics and understanding the impact of these characteristics on the ranking. 
In our case the set of items are the lineups and the output of the model for items $i$ and $j$ provides us essentially with the probability of lineup $\lineup_i$ outperforming $\lineup_j$. 
In particular, 
the Bradley-Terry model is described by \cite{opac-b1127929}:

\begin{equation}
\Pr(\lineup_i \succ \lineup_j | \pi_i,~\pi_j)=\dfrac{e^{\pi_i-\pi_j}}{1+e^{\pi_i-\pi_j}}
\end{equation}
where $\pi_i$ is the {\em ability} of team $i$. 
Given a set of lineup-specific explanatory variables $\mathbf{z}_i$, the difference in the ability of lineups $\lineup_i$ and $\lineup_j$ can be expressed as:

\begin{equation}
\sum_{r=1}^k \alpha_r (z_{ir}-z_{jr}) + U
\end{equation} 
where $U\sim N(0,\sigma^2)$.  
The Bradley-Terry model is then a generalized linear model that can be used to predict the probability of team $i$ winning team $j$.  
In our case, the explanatory variables are the latent space features learned for each lineup, $\lfeature_{\lineup_i}$. 

\begin{figure}[h]
\includegraphics[scale=0.4]{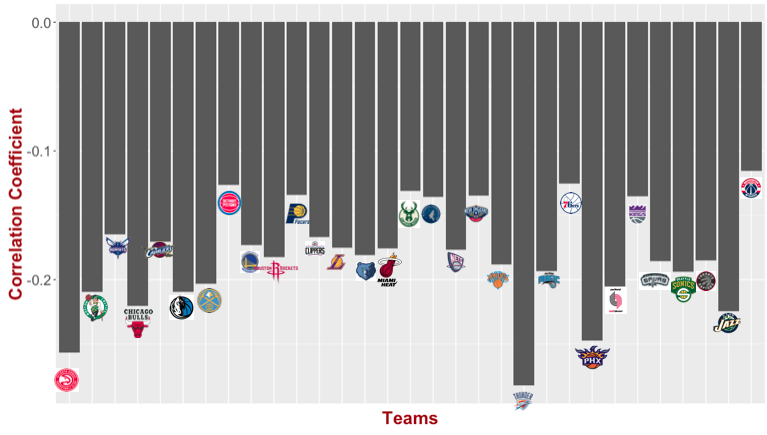}
\caption{{\bf Lineups with higher overlap in terms of players exhibit smaller distance in the latent embedding space $\mathcal{X}$}}
\label{fig:cor}
\end{figure}

{\bf Previously Unseen Lineups: }
One of the challenges (both in out-of-sample evaluations as well as in a real-world setting), is how to treat lineups that we have not seen before, and hence, we do not have their latent space representation. 
In the current design of {\method} we take the following simple approach.  
In particular, for each lineup $\lineup_i$ of team $\mathcal{T}$ we define the similarity  in the players' space $\simil_{\lineup_i,\lineup_j}$ of $\lineup_i$ with $\lineup_j \in \mathcal{L_{\mathcal{T}}}$, as the number of common players between the two lineups.  
It is evident that the similarity value ranges from 0 to 4.  
One might expect that lineups with high overlap in the players' space, should also reside closely in the embedding space.  
In order to get a feeling of whether this is true or not, we calculated for every team and season the correlation between the similarity between two lineups in the players' space (i.e., $\simil_{\lineup_i,\lineup_j}$) with the distance for the corresponding latent features (i.e., ${\tt dist}({\mathbf x}_i,{\mathbf x}_j)$).  
As we can see from Figure \ref{fig:cor} all teams exhibit negative correlations (all correlations are significant at the 0.001 level), which means the more common players two lineups have, the more close they will be projected in the embedding space. 
Of course, the levels of correlation are moderate at best since, the embedding space is obtained by considering the performance of each lineup, and two lineups that differ by only one player might still perform completely differently on the court. 
With this in mind, once we obtain the lineup similarity values, we can assign the latent feature vector for the previously unseen lineup $\lineup_i$ as a weighted average of the lineups in $\mathcal{L_{\mathcal{T}}}$:

\begin{equation}
{\mathbf x}_{\lineup_i} = \dfrac{\displaystyle \sum_{\lineup_j \in \mathcal{L_{\mathcal{T}}}} \simil_{\lineup_i,\lineup_j} \cdot {\mathbf x}_j}{\displaystyle \sum_{\lineup_j \in \mathcal{L_{\mathcal{T}}}} \simil_{\lineup_i,\lineup_j} }
\end{equation}

It is evident that this is simply a heuristic that is currently implemented in {\method}.  
One could think of other ways to approximate the latent space features of a lineup not seen before.  

\subsection{Baselines}
\label{sec:baselines}

For comparison purposes we have also evaluated two baseline approaches for predicting lineup matchup performance. 
The first one is based on network ranking that operates directly on the matchup network (i.e., without involving any embedding of the network), while the second one is based on the adjusted plus/minus rating of the players that belong to the lineup. 

{\bf Network Ranking: }
In our prior work we have shown that ranking teams through a win-loss network, achieves better matchup prediction accuracy as compared to the win-loss percentage \cite{sportsnetrank}. 
Therefore, we follow a similar approach using the lineup matchup network and ranking lineups based on their PageRank score. 
The PageRank of $\net$ is given by: 

\begin{equation}
\bm{r} = D(D-\alpha A)^{-1}\bm{1}
\label{eq:pr}
\end{equation}

\noindent where $A$ is the adjacency matrix of $\net$, $\alpha$ is a parameter (a typical value of which is 0.85) and $D$ is a diagonal matrix where $d_{ii} = \max(1,k_{i,out})$, with $k_{i,out}$ being the out-degree of node $i$.  
Using the PageRank score differential $\Delta r_{ij}=r_{\lineup_i}-r_{\lineup_j}$ as our independent variable we build a logistic regression model for the probability: $\Pr(\lineup_i \succ \lineup_j | \Delta r_{ij})$.

{\bf Adjusted plus/minus (APM): }
The APM statistic of a player is a modern NBA statistic - and for many people the best single statistic we currently have for rating players.  
It captures the additional points that the player is expected to add with his presence in a lineup consisting of league average players matching up with a lineup with league average players. 
APM captures the impact of a player beyond pure scoring.  
For instance, a player might impact the game by performing good screens that lead to open shots, something not captured by current box score statistics.  
The other benefit of APM is that it controls for the rest of the players in the lineups. 
More specifically the APM for a player is calculated through a regression model.  
Let us consider that lineup $\lineup_i$ has played against $\lineup_j$, and has outscored the latter by $y$ points per 48 minutes.  
$y$ is the dependent variable of the model, while the independent variable is a binary vector $\mathbf{p}$, each element of which represents a player. 
All elements of $\mathbf{p}$ are 0 except for the players in the lineups.  
Assuming $\lineup_i$ is the home lineup\footnote{If this information is not available - e.g., because the input data include the total time the lineups matched up over multiple games - W.L.O.G. we can consider the home lineup to be the one with lower ID number.  This is in fact the setting we have in our dataset.}, $p_n = 1,~\forall p_n\in \lineup_i$, while for the visiting lineup, $p_n = -1,~\forall p_n \in \lineup_j$.  
Then these data are used to train a regression model:

\begin{equation}
y = \mathbf{a}^T\cdot \mathbf{p}
\label{eq:apm}
\end{equation}

where $\mathbf{a}$ is the vector of regression coefficients.  
Once obtaining this vector, the APM for player $p_n$ is simply $a_{p_n}$.
The rating of lineup $\lineup_i$, $\rho_{\lineup_i}$ is then the average APM of its players:

\begin{equation}
\rho_{\lineup_i} = \dfrac{a_{p_n}}{5},~\forall p_n \in \lineup_i
\label{eq:lapm}
\end{equation}

Using the lineup rating differential $\Delta \rho_{ij} = \rho_{\lineup_i} - \rho_{\lineup_j}$ as our independent variable we again build a logistic regression model for the probability: $\Pr(\lineup_i \succ \lineup_j | \Delta \rho_{ij})$.

\subsection{Datasets}
\label{data}

In order to evaluate {\method} we used lineup data during the 5 NBA seasons between 2007-08 and 2011-12. 
This dataset includes aggregate information for all the lineup matchups for each of the 5 seasons. 
In particular, for each pair of lineups (e.g., $\lineup_i$, $\lineup_j$) that matched up on the court we obtain the following information:

\begin{enumerate}
\item Total time of matchup
\item Total point differential
\item Players of $\lineup_i$
\item Players of $\lineup_j$
\end{enumerate} 

We used these data in order to obtain both the matchup network as well as to calculate the APM for every player in each season. 

\section{Analysis and Results}
\label{sec:analysis}

We now turn our attention in evaluating {\method}.  
Our focus is on evaluating the accuracy of {\method} in predicting future lineup matchups, as well as the calibration of the inferred probabilities. 
For every season, we build each model using 80\% of the matchups and we evaluate them on the remaining 20\% of the matchups (which might also include lineups not seen before). 
Our evaluation metrics are (i) prediction accuracy, (ii) Brier score and (iii) the probability calibration curves. 

\begin{figure}[h]
\includegraphics[scale=0.4]{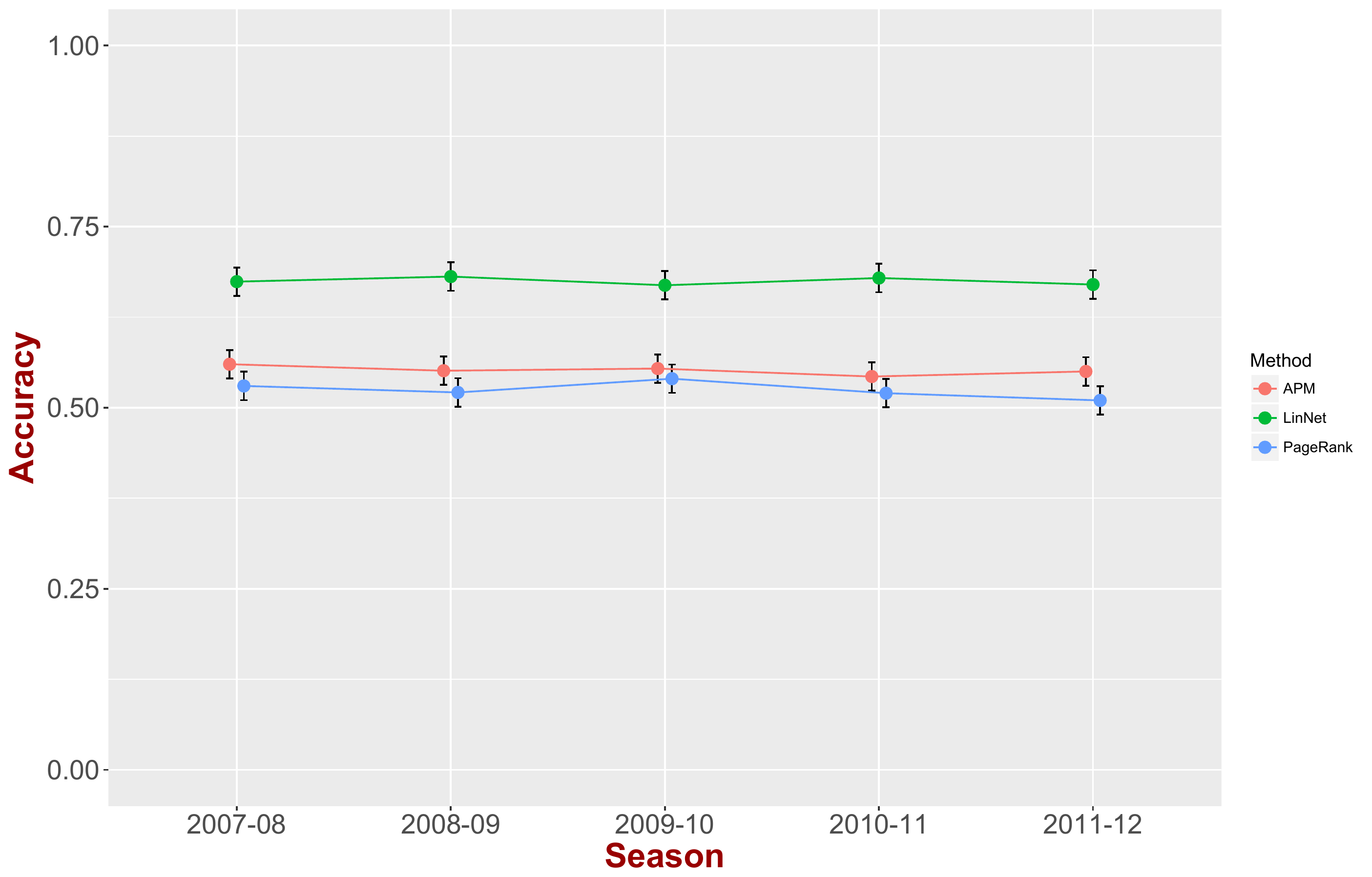}
\caption{{\bf {\method} outperforms in accuracy baseline methods over all 5 seasons examined}}
\label{fig:accuracy}
\end{figure}

Figure \ref{fig:accuracy} presents the accuracy performance of the various methods. 
As we can see {\method} outperforms both the PageRank and APM systems over all 5 seasons examined. 
{\method}'s accuracy is 67\%, while APM exhibits a 55\% average accuracy and PageRank a 53\% accuracy. 

However, equally as important for the quality of the model is the calibration of the output probabilities. 
We begin by first computing the Brier score \cite{brier1950verification} for each model and dataset. 
In the case of a binary probabilistic prediction the Brier score is calculated as: 

\begin{equation}
\brier = \dfrac{1}{N}\sum_{i=1}^N (\pi_i-y_i)^2
\label{eq:brier}
\end{equation}
where $N$ is the number of observations, $\pi_i$ is the probability assigned to instance $i$ being equal to 1 and $y_i$ is the actual (binary) value of instance $i$.  
The Brier score takes values between 0 and 1 and evaluates the calibration of these probabilities, that is, the level of confidence they provide (e.g., a 0.9 probability is {\em better} calibrated compared to a 0.55 probability when the ground truth is label 1).   
The lower the value of $\brier$ the better the model performs in terms of calibrated predictions. 
Our model exhibits an average Brier score $\brier$ of 0.19, while both PageRank and APM models have a worse Brier score.  
Typically the Brier score of a model is compared to a baseline value $\brier_{base}$ obtained from a {\em climatology} model \cite{mason2004using}. 
A climatology model assigns the same probability to every observation, which is equal to the fraction of positive labels in the whole dataset.  
Therefore, in our case the climatology model assigns a probability of 0.5 to each observation. 
As alluded to above we do not have information about home and visiting lineup so our model estimates the probability of the lineup with the smaller ID outperforming the one with the larger ID.  
Given that the lineup ID has no impact on this probability the climatology model probability is 0.5.  
The Brier score for this reference model is $\brier_{base}=0.25$, which is of lower quality as compared to {\method} and also slightly worse than our baselines.

\begin{figure}[h]
\includegraphics[scale=0.4]{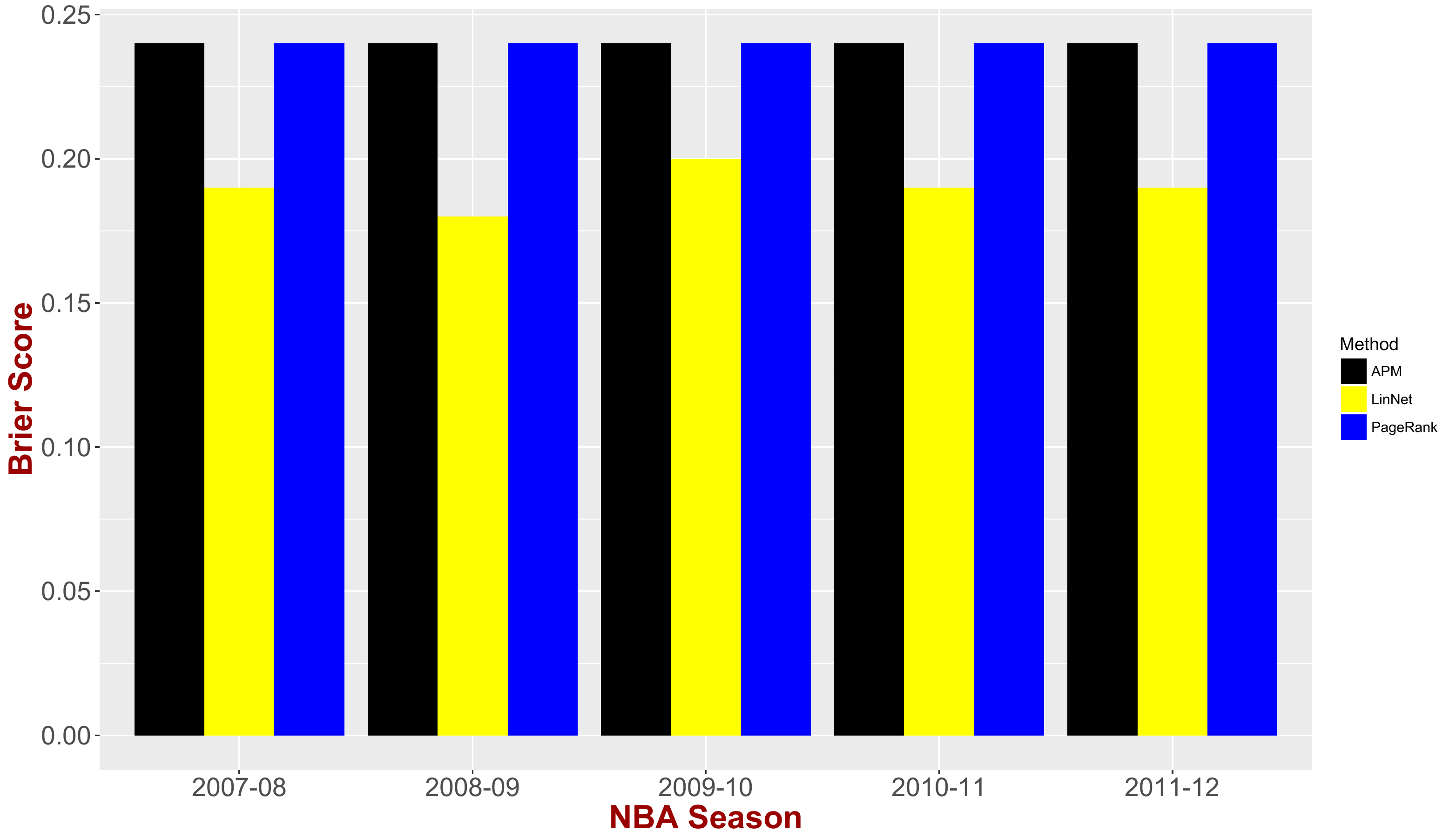}
\caption{{\bf {\method} exhibits better calibrated probabilities as compared to the baselines (smaller Brier score translates to better calibration)}}
\label{fig:brier}
\end{figure}

\begin{figure}[h]
\includegraphics[scale=0.4]{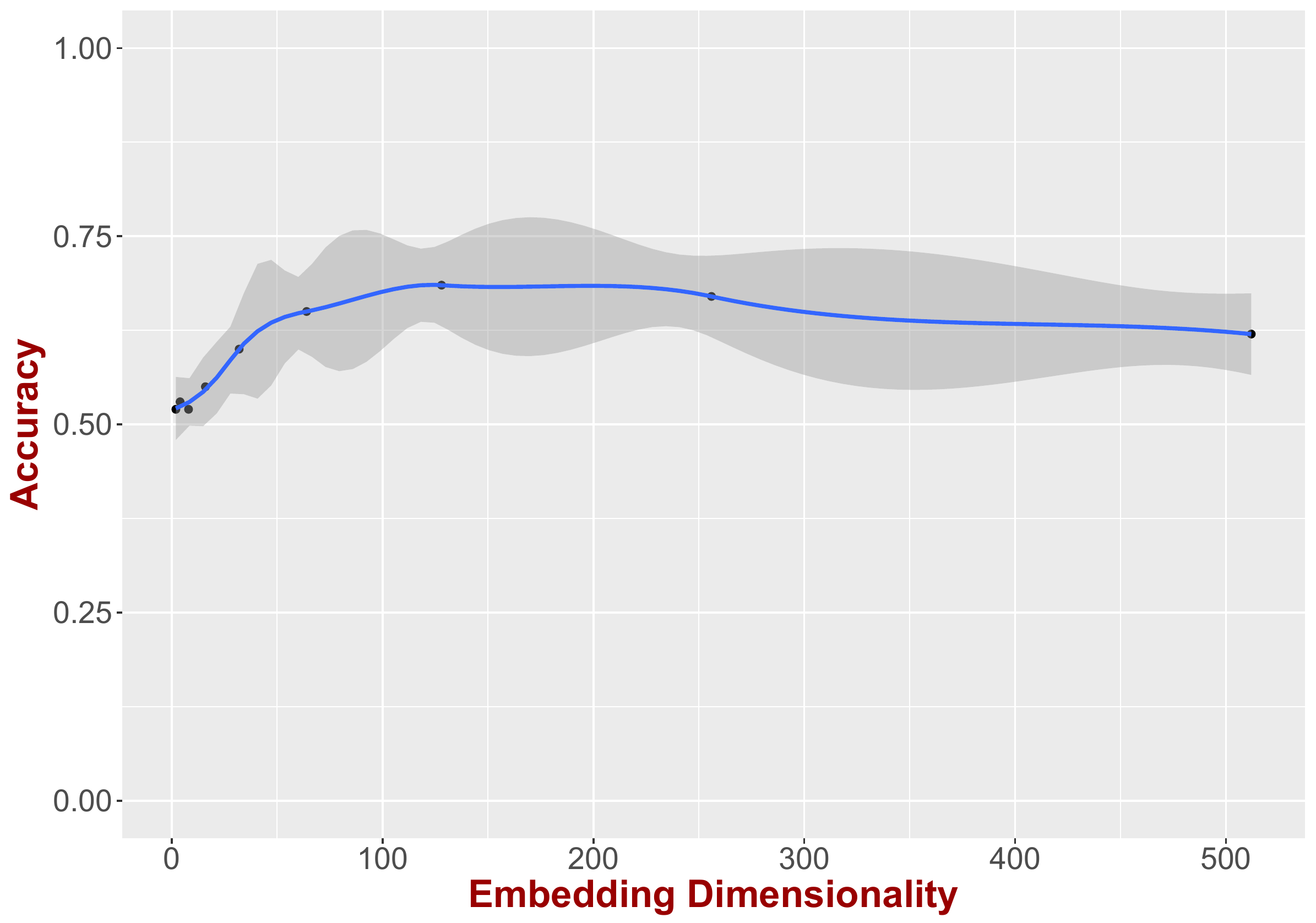}
\caption{{\bf The choice of $d=128$ for the embedding dimensionality of {\method} provides a good tradeoff between accuracy and (computational) complexity}}
\label{fig:accuracy-d}
\end{figure}

As alluded to above we have picked a dimensionality for the embedding of $d=128$. 
However, we have experimented with different embedding dimensionality values and our results are presented in Figure \ref{fig:accuracy-d}.  
As we can see, low dimensionality does not provide any benefit over the baselines, while increasing further the dimensionality (above 128) exhibits diminishing returns. 

\begin{figure}[h]
\includegraphics[scale=0.4]{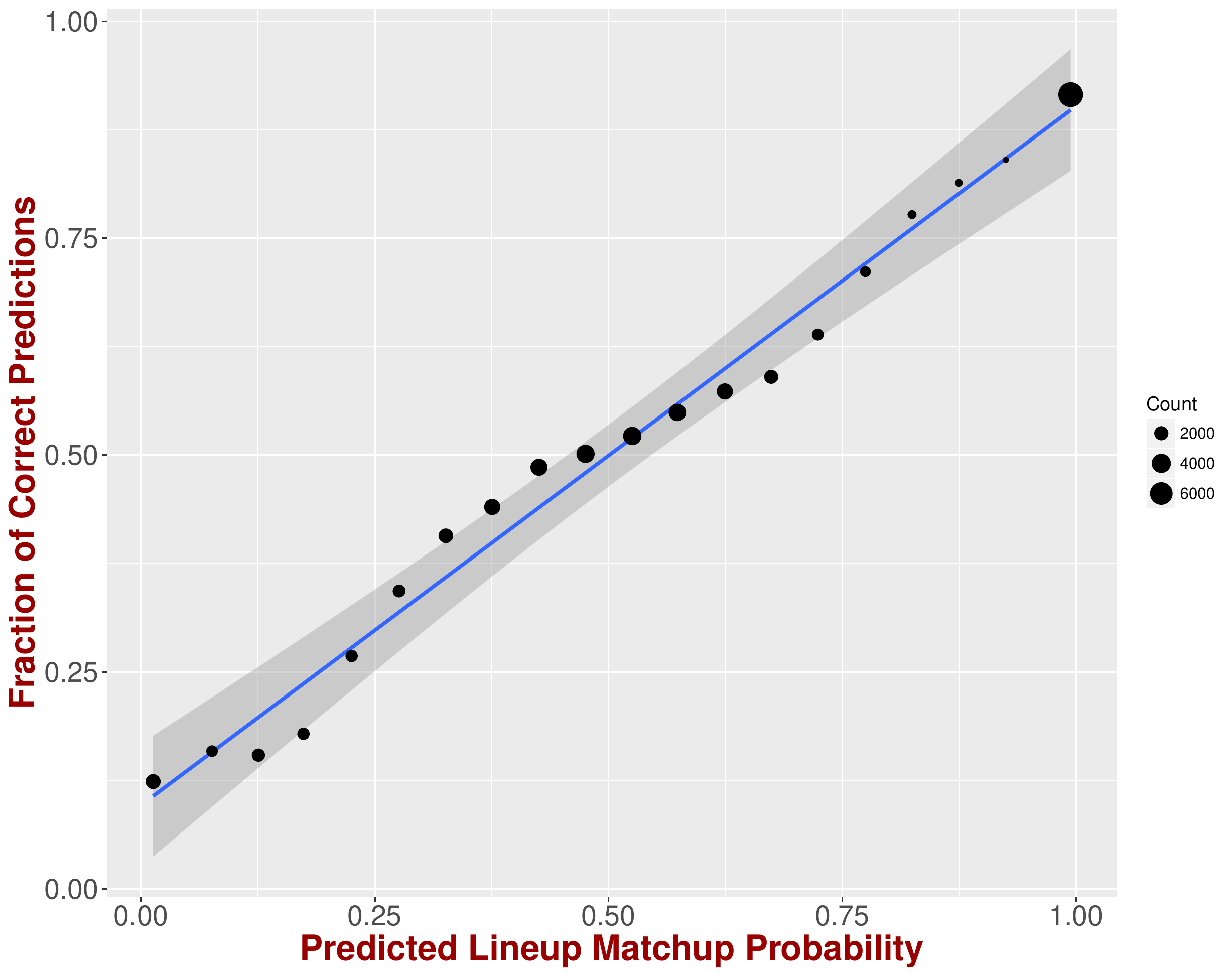}
\caption{{\bf The {\method} probability validation curve is very close to the $y=x$ line, translating to fairly accurate probability estimations}}
\label{fig:calibration}
\end{figure}

Finally, we evaluate the accuracy of the probability output of {\method} by deriving the probability validation curves (for $d=128$). 
In order to evaluate this we would ideally want to have every matchup played several times. 
If the favorite lineup were given a 75\% probability of outperforming the opposing lineup, then if the matchup was played 100 times we would expect the favorite to outperform in 75 of them. 
However, this is not realistic and hence, in order to evaluate the accuracy of the probabilities we will use all the games in our dataset. 
In particular, if the predicted probabilities were accurate, when considering all the matchups where the favorite was predicted to win with a probability of x\%, then the favorite should have outperform the opponent in x\% of these matchups. 
Given the continuous nature of the probabilities we quantize them into groups that cover a 5\% probability range. 
Fig \ref{fig:calibration} presents the predicted win probability for the reference lineup (i.e., the lineup with the smaller ID) on the x-axis, while the y-axis presents how many of these matchups this reference lineup won. 
Furthermore, the size of the points represents the number of instances in each situation. 
As we can see the validation curve is very close to the $y=x$ line, which practically means that the predicted probabilities capture fairly well the actual matchup probabilities. 
In particular, the linear fit has an intercept of 0.1 and a slope of 0.85. 

\begin{figure}[h]
\includegraphics[scale=0.4]{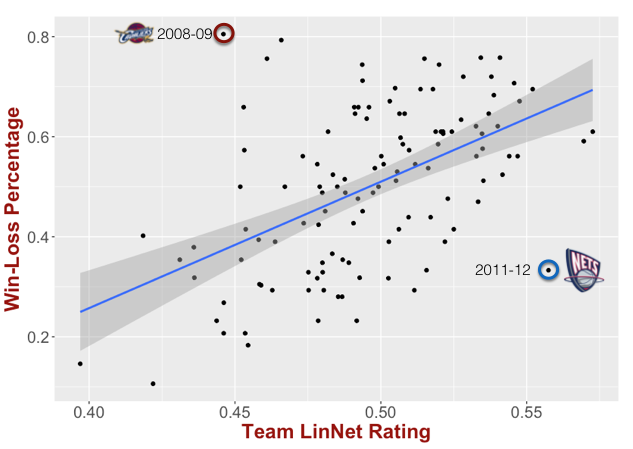}
\caption{{\bf The {\method} probability validation curve is very close to the $y=x$ line, translating to fairly accurate probability estimations}}
\label{fig:lineup-wl}
\end{figure}

{\bf Season Win-Loss Percentage and Lineup Performance: }
How well can lineup {\em ratings} obtained from {\method} explain the win-loss record of a team? 
One should expect that there is a correlation between {\method} lineup ratings and the record of a team - which as we will see indeed is the case.  
However, this correlation is also not expected to be perfect, since it relies also on coaching decisions as well as availability of the lineups (e.g., a lineup can be unavailable due to injuries). 
In order to examine this we focus on lineups that played for a total of more than a game (i.e., 48 minutes) during the season. 
Then with $p_{\lineup_i}$ being the average probability of lineup $\lineup_i$ (of team $\tau$) outperforming each of the opponent's lineups (i.e., $p_{\lineup_i} = \dfrac{\sum_{\lineup_j \in \mathcal{L}\setminus\mathcal{L}_{\tau}} \Pr(\lineup_i \succ \lineup_j)}{|\mathcal{L}\setminus\mathcal{L}_{\tau}|}$, where $\mathcal{L}_{\tau}$ is the set of all lineups of team $\tau$ and $\mathcal{L}$ is the set of all league lineups), the {\method} team rating of team $\tau$ is: 

\begin{equation}
\rating(\tau) = \dfrac{\displaystyle\sum_{\lineup_i \in \mathcal{L}_{\tau}} \gamma_i\cdot p_{\lineup_i}}{\displaystyle\sum_{\lineup_i \in \mathcal{L}_{\tau}} \gamma_i}
\label{eq:team-rating}
\end{equation}
where $\gamma_i$ is the total time lineup $\lineup_i$ has been on the court over the whole season. 
Our results are presented in Figure \ref{fig:lineup-wl}.  
The linear regression fit has a statistically significant slope (p-value $<$ 0.001), which translates to a statistically important relationship.  
However, as we can see there are outliers in this relationship, such as the 2008-09 Cavaliers and the 2011-12 Nets.  
The linear relationship explains 27\% of the variability at the win-loss records of the teams. 
This might be either because teams do not choose (due to various reasons) their best lineup to matchup with the opponent, or because the time that a lineup is on court is important for its performance (we discuss this in the following section), something that {\method} currently does not account for.  
Overall, the correlation coefficient between the {\method} team rating and the win-loss record is 0.53 (p-value $<$ 0.0001).   

\section{Discussion}
\label{sec:discussion}

In this work we presented {\method}, a network embedding approach in evaluating lineups. 
Our evaluations indicate that the probability output from {\method} is well calibrated and more accurate than traditional lineup evaluation methods. 
However, there are still some open issues with the design of {\method}.  
More specifically, a matchups between specific lineups might last only for a few minutes (or even just a couple of possessions).  
This creates a reliability issue with any predictions one tries to perform with similar information.  
Even though we adjust the performance margin on a per minute basis, it is not clear that a lineup can keep up its performance over a larger time span. 
Furthermore, currently for lineups we have not seen before we use as its latent features a weighted average of already seen lineups of the team, weighted based on their similarity in the players' space. 
However, there are other approaches that one might use for this task that could potentially provide even better results. 
For example, a regression (similar to the adjusted plus/minus) can be used to infer the latent features based on the players in the lineup. 
This is something that we plan in exploring in the future. 

\small
\bibliography{main-abbrv}


\end{document}